\documentclass[twocolumn,superscriptaddress,floatfix,preprintnumbers, nofootinbib,hyperref]
{revtex4-2} 
\pdfoutput=1
\usepackage[colorlinks=true,breaklinks=true]{hyperref}
\usepackage{float}
\usepackage[normalem]{ulem}
\usepackage[utf8]{inputenc}
\hypersetup{allcolors=[rgb]{0.0 0.0 0.6},linkcolor=[rgb]{0.75 0.05 0.05}}
\usepackage{amsmath,amssymb} \usepackage{epsfig}  
\usepackage{graphicx}   
\usepackage{color}
\usepackage{multirow}
\usepackage{comment}
\usepackage{amssymb}
\usepackage{orcidlink}
\usepackage{soul}
\usepackage[version=3]{mhchem}
\DeclareUnicodeCharacter{2212}{-}

\allowdisplaybreaks

\bibpunct{[}{]}{,}{n}{}{,}
\definecolor{ForestGreen}{RGB}{34,139,34}

\begin{document}

\title{Axion-photon conversion down to the nonrelativistic regime}

\author{Clemente~Smarra \orcidlink{0000-0002-0817-2830}}
\email{csmarra@sissa.it}
\affiliation{SISSA International School for Advanced Studies, Via Bonomea 265, 34136, Trieste, Italy}
\affiliation{INFN,  Sezione di Trieste, Via Bonomea 265, 34136, Trieste, Italy}

\author{Pierluca~Carenza}\email{pierluca.carenza@fysik.su.se}
\affiliation{The Oskar Klein Centre, Department of Physics, Stockholm University, Stockholm 106 91, Sweden
}

\smallskip

\begin{abstract}
\noindent
In the presence of a magnetic field, axions can convert into photons and vice versa. The phenomenology of the conversion is captured by a system of two coupled Klein-Gordon equations, which, assuming that the axion is relativistic, is usually recast into a pair of first-order Schr\"odinger-like equations. In such a limit, focusing on a constant magnetic field and plasma frequency, the equations admit an exact analytic solution. The relativistic limit significantly simplifies the calculations and, therefore, it is widely used in phenomenological applications.
In this work, we discuss how to evaluate the axion-photon system evolution without relying on such relativistic approximation. In particular, we give an exact analytical solution, valid for any axion energy, in the case that both the magnetic field and plasma frequency are constant.
Moreover, we devise an analytic perturbative expansion that allows for tracking the conversion probability in a slightly inhomogeneous magnetic field or plasma frequency, whose characteristic scale of variation is much larger than the typical axion-photon oscillation length. Finally, we discuss a case of resonant axion-photon conversion giving useful simplified formulae that might be directly applied to dark matter axions converting in neutron star magnetospheres.
\end{abstract}

\maketitle

\section{Introduction}
The nonobservation of \textit{CP} violation in strong interactions, commonly dubbed as strong-\textit{CP} problem,  still stands as one of the unsolved puzzles of the standard model.
One of the simplest and most compelling solutions is represented by the Peccei-Quinn mechanism, which implies the existence of a hypothetical pseudoscalar particle called the axion~\cite{Peccei:1977ur,Peccei:1977hh,Weinberg:1977ma,Wilczek:1977pj}.
Several axion models were proposed, starting from the ones were new physics was appearing at the electroweak scale~\cite{Weinberg:1977ma,Wilczek:1977pj}, soon ruled out by experiments, to invisible axion models~\cite{Kim:1979if,Shifman:1979if,osti_7063072,Dine:1981rt} (see~\cite{DiLuzio:2020wdo} for a review). Currently, invisible axion models are not in contrast with observations and most of the parameter space is still unprobed. A common feature of axion models is an axion-photon interaction described by the following Lagrangian
\begin{equation}
    \mathcal{L}\supset-\frac{1}{4}g_{a\gamma}a\,F_{\mu\nu}\tilde{F}^{\mu\nu}\,,
    \label{eq:lag}
\end{equation}
where $a$ is the axion field, $F_{\mu\nu} = \partial_\mu A_\nu - \partial_\nu A_\mu$ is the electromagnetic field strength tensor, $\tilde{F}^{\mu \nu}$ is its dual and $g_{a\gamma}$ is the axion-photon coupling. 
Furthermore, many grand unified theories and string theory realizations predict the existence of pseudoscalars interacting with photons via the interaction in Eq.~\eqref{eq:lag}~\cite{Green:1987sp,Svrcek:2006yi,Halverson:2019cmy,Demirtas:2021gsq}. 
Thus, it is of primary importance to explore the phenomenology of hypothetical light pseudoscalar particles interacting with photons. This research line is extremely active and, nowadays, it is attracting a huge theoretical and experimental effort.

The axion-photon vertex in Eq.~\eqref{eq:lag} enables axions to convert into photons, and vice versa, in an external electromagnetic field. This phenomenon is widely used to produce and detect axions in laboratory experiments~\cite{Anselm:1985obz,VanBibber:1987rq, Ehret:2010mh, Bahre:2013ywa}, reveal axions that constitute the dark matter (DM)~\cite{Sikivie:1983ip,Asztalos:2003px,Brun:2019lyf} or probe axions produced in the Sun~\cite{Sikivie:1983ip,vanBibber:1988ge,Armengaud:2019uso}. 
There have been several theoretical studies on the formalism of axion-photon conversions, e.g.~\cite{MAIANI1986359, Sikivie:1983, Sikivie:1985, 
Raffelt:1987im,
Pshirkov:2007st, 
Laming_2013,
Huang_2018, Arza_2019, 
Leroy:2019ghm,
Battye:2019aco, Millar:2021gzs, Carenza:2023nck,McDonald:2023ohd,McDonald:2024nxj}. In this paper, we focus on the formalism presented in Ref.~\cite{Raffelt:1987im}, generalizing the axion-photon mixing treatment to cases in which axions are not relativistic, as usually assumed. 
This generalization is of interest for the phenomenology of DM axions converting into astrophysical magnetic fields, for example in the surroundings of a neutron star~\cite{Battye:2019aco,Leroy:2019ghm,Witte:2021arp,Battye:2021xvt,Witte:2022cjj,McDonald:2023shx,Tjemsland:2023vvc}, where axions are accelerated to mildly relativistic speeds, at most, by the neutron star gravitational attraction before converting into photons. See also Ref.~\cite{Ai:2024cja} for a related discussion.

This paper is structured as follows. Taking both the magnetic field and the plasma frequency to be constant, in Sec.~\ref{sec:basics} we discuss the traditional approach to axion-photon oscillations, closely following Ref.~\cite{Raffelt:1987im}, in order to highlight which assumptions are made. In Sec.~\ref{sec:exact}, we show how to generalize this calculation by dropping the assumption of relativistic axions, leading to a generalized equation for the axion-photon mixing. In particular, we give an exact solution for the axion-photon mixing in a constant magnetic field and plasma frequency, showing how to recover the well-known results in the relativistic limit. In Sec.~\ref{sec:pert}, we discuss how to calculate the conversion probability in the case of an almost-homogeneous magnetic field or plasma frequency, developing a perturbative approach valid for small inhomogeneities. In Sec.~\ref{sec:LZ}, we show the discrepancies between the relativistic approximation and the exact solution in the case of resonant conversion, providing a modification of the Landau-Zener formula that correctly reproduces the conversion probability for any axion speed.
In Sec.~\ref{sec:concl}, we summarize and conclude the paper. Finally, in the Appendix, we review different formalisms used in literature to study the axion-photon system dynamics, comparing them with our approach and highlighting similarities and differences.

\section{Axion-photon mixing basics}
\label{sec:basics}

When an external magnetic field $\vec{B}$ is present, axions and photons can convert into each other. The axion-photon system is described by the following Lagrangian~\cite{Raffelt:1987im}:
\begin{equation}
\begin{split}
    \mathcal{L}=& -\frac{1}{4}F_{\mu\nu}F^{\mu\nu} - \frac{1}{4}g_{a\gamma}a\,F_{\mu\nu}\tilde{F}^{\mu\nu}+\frac{1}{2}\partial_\mu a\partial^\mu a -\frac{m_a^2}{2} a^2 + \\
    &+\frac{\alpha^2}{90 m_e^4}\left[\left(F_{\mu\nu}F^{\mu\nu}\right)^2 +\frac{7}{4}\left(F_{\mu\nu}\tilde{F}^{\mu\nu}\right)^2 \right]\,,
\end{split}
    \label{eq:lagr}
\end{equation}
where the axion field has a mass $m_{a}$ and a coupling to photon $g_{a\gamma}$, $\alpha$ is the fine-structure constant and $m_{e}$ is the electron mass.
We restrict to propagation along the $z$-axis of the axion-photon beam. This simplification makes the analytical treatment that follows possible, however, it might miss some effects that emerge in a pure three-dimensional approach~\cite{McDonald:2023ohd}.
Under these assumptions, the equations of motion of the axion-photon system are
\begin{equation}
\begin{split}
  & \left[\Box - 
   \mathcal{M}^{2}\right]
   \begin{pmatrix}
        A_\perp(z,t)\\
        A_\parallel(z,t)\\
        a(z,t)
    \end{pmatrix} = 0\,,\\
    &\mathcal{M}^{2} = \begin{pmatrix}
        Q_\perp&0&0\\
        0&Q_\parallel&Q_{a\gamma}\\
        0&Q_{a\gamma}&Q_a\\
    \end{pmatrix}
    \label{eq:eom},
    \end{split}
\end{equation}
where $A_\parallel$ and $A_\perp$ are the photon field components parallel and perpendicular to the external magnetic field $ \vec{B}$. We consider a nonoptically active medium, neglecting possible Faraday rotation effects. Focusing on stationary solutions, namely $F(z, t) \propto e^{i\omega t}F(z)$, where $F = {A_\perp, A_\parallel, a}$, we can thus write
\begin{equation}
    \begin{split}
        Q_{i}&=2\omega^{2}(n_{i}-1)-\omega_{\rm pl}^{2}\,,\quad{\rm for}\quad i=\{\perp,\parallel\}\,,\\
        Q_{a\gamma}&=g_{a\gamma}\omega\,B_{0}\,,\\
        Q_{a}&=-m_{a}^{2}\,,\\
        n_{\perp}&=1+2\zeta\sin^{2}\Theta\,,\\
        n_{\parallel}&=1+\frac{7}{2}\zeta\sin^{2}\Theta\,,\\
        \zeta&=\frac{\alpha}{45\pi}\left(\frac{B_{0}}{B_{\rm crit}}\right)^{2}\,,
    \end{split}
    \label{eq:invac}
\end{equation}
where $\omega$ is the axion or photon energy, $\omega_{\rm pl}$ is the plasma frequency,\footnote{In a cold, nonrelativistic medium, the plasma frequency is $\omega_\text{pl}^{2} = 4\pi\alpha n_e/m_e$, where $n_e$ is the electron density. We restrict to this specific form of $\omega_\text{pl}$, although a more careful treatment is needed in extreme environments, like neutron star magnetospheres~\cite{McDonald:2023ohd}.} $B_{\rm crit}=4.41\times10^{13}$~G and $\Theta$ is the angle between the magnetic field vector, with intensity $B_0 = \lvert\vec{B}\rvert$, and the photon propagation direction.\footnote{ For a given energy $\omega$ and axion-photon coupling $g_{a\gamma}$, we focus on sufficiently low values of the magnetic field $B_0$, such that $n_i$ is correctly described by the form in Eq.~\eqref{eq:invac}. See Ref.~\cite{Fairbairn_2011} for further details.} Here, for simplicity, we neglect dispersion effects on the cosmic microwave background, relevant above TeV energies~\cite{Dobrynina:2014qba}.

Having neglected Faraday rotation effects, we can then decouple $A_{\perp}$ from the evolution and just focus on the $A_\parallel -a$ system. Therefore, we can rewrite Eq.~\eqref{eq:eom} as
\begin{equation}
    \begin{split}
        &\left[\omega^{2}+\partial_{z}^{2}+\mathcal{M}^{2}\right] \Psi = 0\,,\\
            &\Psi=\begin{pmatrix}
        A_\parallel(z)\\
        a(z)
    \end{pmatrix}\,.
    \end{split}
    \label{eq:simpleom}
\end{equation}
Now, we assume that the mass matrix in Eq.~\eqref{eq:simpleom} does not depend on $z$. In other words, we are assuming both a constant magnetic field, with modulus $\lvert\vec{B}\rvert = B_{0}$ and fixed direction, and that conversion takes place in a homogeneous plasma, for which also the plasma frequency is constant. In this case, we can diagonalize the mass matrix and go to the unmixed basis, where Eq.~\eqref{eq:simpleom} reads

\begin{align}
    &\left[\omega^2 + \partial_z^2 + 
    \mathcal{M}_\text{diag}^2\right]
    \Psi' = 0,\label{eq:eomdiag}\\
    & \Psi' \equiv \begin{pmatrix}
        A'_\parallel(z)\\
        a'(z)
    \end{pmatrix} = \mathcal{O}
    \begin{pmatrix}
        A_\parallel(z)\\
        a(z)
    \end{pmatrix} \nonumber\\
    &\quad\equiv  \begin{pmatrix}
        \cos\theta&\sin\theta\\
        -\sin\theta&\cos\theta\\
    \end{pmatrix}\begin{pmatrix}
        A_\parallel(z)\\
        a(z)
    \end{pmatrix}, \label{eq:thetamat}\\
    &\mathcal{M}^2_\text{diag} =  \begin{pmatrix}
        Q'_\parallel&0\\
        0&Q'_a\\
    \end{pmatrix},\label{eq:m2diagonal}
\end{align}
with 
\begin{align}
    &\frac{1}{2}\tan2\theta = \frac{Q_{a\gamma}}{Q_\parallel -Q_a},\label{eq:tantheta}\\
    &Q'_\parallel = \frac{Q_\parallel + Q_a}{2}+ \frac{Q_\parallel - Q_a}{2\cos2\theta} = \frac{1}{2}\left(Q_\parallel + Q_a + \Delta Q \right),\\
    &Q'_a = \frac{Q_\parallel + Q_a}{2}- \frac{Q_\parallel - Q_a}{2\cos2\theta} = \frac{1}{2}\left(Q_\parallel + Q_a -\Delta Q\right),
\end{align}
where we have defined, for simplicity,
\begin{equation}
    \Delta Q \equiv \sqrt{\left(Q_\parallel - Q_a\right)^2 + 4Q_{a\gamma}^2}.
    \label{eq:dq}
\end{equation}
It is easy to notice that, in the diagonal basis, Eq.~\eqref{eq:eomdiag} describes a system of two decoupled equations. In order to solve them, we can impose the following ansatz on the rotated fields
\begin{equation}
\begin{pmatrix}
        A'_\parallel(z)\\
        a'(z)
\end{pmatrix} = \begin{pmatrix}
    A'_\parallel\, e^{-ik_\parallel' z}\\
    a'\, e^{-ik_a' z}
\end{pmatrix} + \text{c.c.}, 
\label{eq:pw}
\end{equation}
where we assume that the coefficients $A'_\parallel, a'$ do not depend on $z$ or, more precisely, that $\partial_z A_\parallel', \partial_z a' \ll k$. We will avoid writing the complex conjugate ($\text{c.c.}$) part of the solution in the following computations.
Therefore, in the relativistic limit, we can reduce Eq.~\eqref{eq:eomdiag} to a first-order equation, using the approximation
\begin{equation}
\begin{split}
    \omega^{2}+\partial_{z}^{2}&= (\omega-i \partial_{z})(\omega+i \partial_{z})=\\
        &= (\omega-i \partial_{z})(\omega+k)\simeq2\omega(\omega-i \partial_{z})
        \end{split}\,,
    \label{eq:dalrel}
\end{equation} 
where $k$ is a placeholder for $k_\parallel', k_a'$. Then, Eq.~\eqref{eq:eomdiag} becomes
\begin{equation}
    \left[\omega - i\partial_z+ \frac{1}{2\omega} 
    \mathcal{M}_\text{diag}^2\right] \Psi ' = 0,
    \label{eq:t1before}
\end{equation}
which, upon a phase redefinition of $\Psi'$,\footnote{As we will in general deal with conversion probabilities, common phases can be simply neglected in the analysis.} leads to the usual Schr\"odinger-like equation~\cite{Raffelt:1987im}
\begin{equation}
    i\partial_{z}\Psi'= \frac{1}{2\omega}\mathcal{M}_\text{diag}^{2} \Psi' \,.
\label{eq:t1}
\end{equation}
Let us point out a subtlety that is often overlooked in the literature. In order to reduce the system in Eq.~\eqref{eq:simpleom} to a first-order Schr\"odinger-like form, it is common to apply the condition in Eq.~\eqref{eq:dalrel} directly to the mixed system Eq.~\eqref{eq:simpleom}. This amounts to having Eq.~\eqref{eq:t1} expressed in terms of the vector $\Psi$ instead of $\Psi'$, with $\mathcal{M}^2$ replacing $\mathcal{M}^2_\text{diag}$. However, as we showed before, the condition in Eq.~\eqref{eq:dalrel} is strictly valid when imposing a plane wave condition on the fields, as we consistently do in Eq.~\eqref{eq:pw}. Such a requirement can only be meaningful for eigenstates, because these are the conserved quantities in the motion. Fields that are not eigenstates, such as the ones described by $\Psi$ in Eq.~\eqref{eq:simpleom}, continuously convert into each other, and so their behavior cannot be modeled by Eq.~\eqref{eq:pw}.\footnote{When the axion is ultrarelativistic, however, Eq.~\eqref{eq:t1before} implies that both of the eigenstates, and thus any of their combinations, oscillate as $\sim e^{i\omega z}$, as we can simply neglect the mass terms. This justifies what it is usually done, which is anyway inconsistent from a conceptual standpoint.} In conclusion, to be consistent, we should first rotate the second-order Eq.~\eqref{eq:simpleom} to the diagonal basis, and then we can impose the condition in Eq.~\eqref{eq:dalrel} to reduce it to the first-order system in Eq.~\eqref{eq:t1}.

Plugging the ansatz Eq.~\eqref{eq:pw} in Eq.~\eqref{eq:t1}, valid in the relativistic limit, and using Eq.~\eqref{eq:m2diagonal} yields
\begin{equation}
    \begin{pmatrix}
        k_\parallel'\\
        k_a'
    \end{pmatrix} = \frac{1}{2\omega}
    \begin{pmatrix}
        Q_\parallel'\\
        Q_a'
    \end{pmatrix}\equiv 
    \begin{pmatrix}
        \Delta_\parallel'\\
        \Delta_a'
    \end{pmatrix}
    \label{eq:ks}
\end{equation}
and, thus,
\begin{equation}
    \Psi' = \begin{pmatrix}
        A_\parallel'\,e^{-i\Delta'_\parallel z }\\
        a'\,e^{-i\Delta'_a z}
    \end{pmatrix} = \begin{pmatrix}
        e^{-i\Delta'_\parallel z }&0\\
        0&e^{-i\Delta'_a z }
    \end{pmatrix} 
    \begin{pmatrix}
        A_\parallel'\\
        a'
    \end{pmatrix},
    \label{eq:solprime}
\end{equation}
where $A_\parallel'$ and $a'$ are the wave functions of the propagation eigenstates at the initial point $z=0$. However, experiments detect axions and photons in their interaction eigenstates, not in the propagation ones. Therefore, it would be useful both to impose initial conditions and to have final results in the original mixed basis of Eq.~\eqref{eq:simpleom}. From Eq.~\eqref{eq:solprime}, we immediately see that
\begin{equation}
\begin{split}
    \begin{pmatrix}
        A_\parallel(z)\\
        a(z)
    \end{pmatrix} &= \mathcal{O}^{-1}\begin{pmatrix}
        e^{-i\Delta'_\parallel z }&0\\
        0&e^{-i\Delta'_a z }
    \end{pmatrix} \mathcal{O}     \begin{pmatrix}
        A_\parallel\\
        a
    \end{pmatrix}\equiv\\
    &\equiv M(z)\begin{pmatrix}
        A_\parallel\\
        a
    \end{pmatrix},
\end{split}
\label{eq:evol}
\end{equation}
with $\mathcal{O}$ defined in Eq.~\eqref{eq:thetamat} and $\{A_\parallel, a\}$ labeling the initial abundance of photons and axions. The matrix $M(z)$ encodes the evolution of the axion-photon system from the initial point $z = 0$ up to a generic $z$. We stress that, in order to determine this solution, we used the equation of motion in Eq.~\eqref{eq:t1}, which is valid in the relativistic limit. Therefore, the following results are valid only in this regime.

The axion-photon conversion probability, given by the square of the absolute value of the off-diagonal terms of the evolution matrix $M$, reads
\begin{equation}
\begin{split}
    P_\text{rel}(z) &= \lvert M_\text{12}(z) \rvert^2 = \left|\cos\theta\sin\theta\left[e^{-i\Delta_\parallel' z} - e^{-i\Delta_a'z}\right]\right|^2\\
    &=\frac{4\Delta_{a\gamma}^2}{\Delta_\text{osc}^2}\sin^2\left(\frac{\Delta_\text{osc}}{2}z\right),
\end{split}
\label{eq:probrel}
\end{equation}
with
\begin{align}
    &\Delta_\text{osc} = \sqrt{\left(\Delta_\parallel - \Delta_a\right)^2 + 4\Delta_{a\gamma}^2},
    \label{eq:doscrel}
\end{align}
where, analogously to Eq.~\eqref{eq:ks}, we defined
\begin{equation}
    \Delta_{i} \equiv \frac{Q_{i}}{2\omega}\qquad i \in \{\parallel, a, a\gamma\}.\\
    \label{eq:di}
\end{equation}
The subscript ``rel" in Eq.~\eqref{eq:probrel} reminds us that this conversion probability is valid uniquely in the relativistic limit. As a side note, Eq.~\eqref{eq:evol} also informs us about the right prescription to use to determine the initial conditions on the axion and photon field derivatives if we want to tackle the problem numerically. For instance, in a scenario where there is one axion and no photons, assuming $a(0) = 1, \partial_z a(z)|_{z=0} = 0$ as a pair of the initial condition is inconsistent. Indeed, by expanding the evolution matrix $M$ in Eq.~\eqref{eq:evol}, we see that the first derivative reads
\begin{equation}
    \left.\partial_z a(z)\right|_{z=0} = -i\left(\Delta'_a \cos^2\theta + \Delta'_\parallel \sin^2\theta\right),
\end{equation}
which is, in general, different from 0.
In the following sections, we present frameworks in which the resulting evolution matrix, describing the axion-photon conversion, is generally different from Eq.~\eqref{eq:evol}. Provided we replace the relativistic evolution matrix $M$ with the appropriate one for the case under examination, the same considerations apply.

\section{Exact solution of axion-photon mixing}
\label{sec:exact}
\begin{figure}[t!]
        \includegraphics[width=0.9\columnwidth]{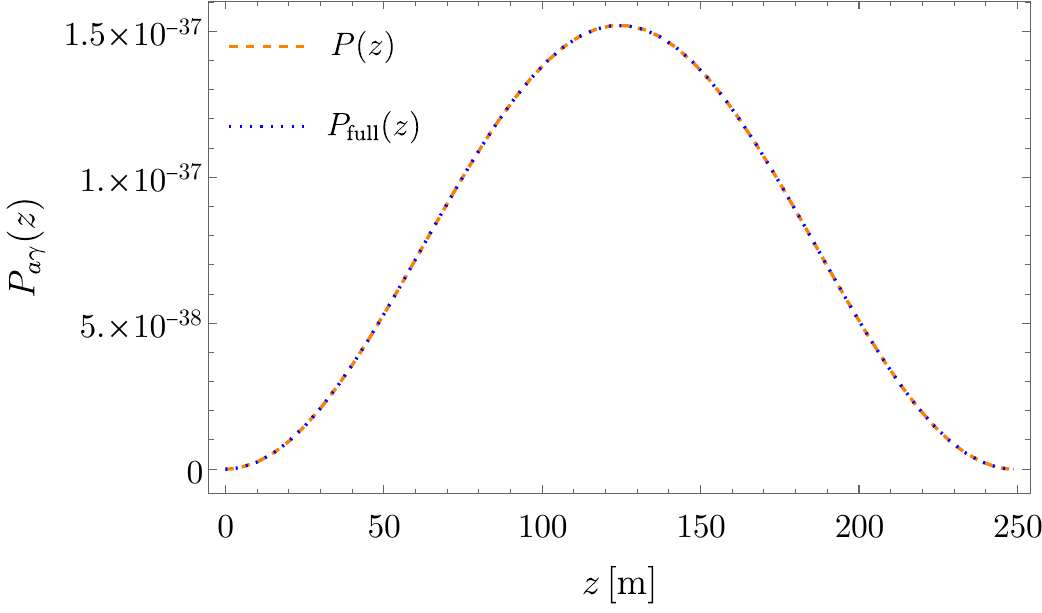}
        \label{fig:comp}
 \includegraphics[width=0.9\columnwidth]{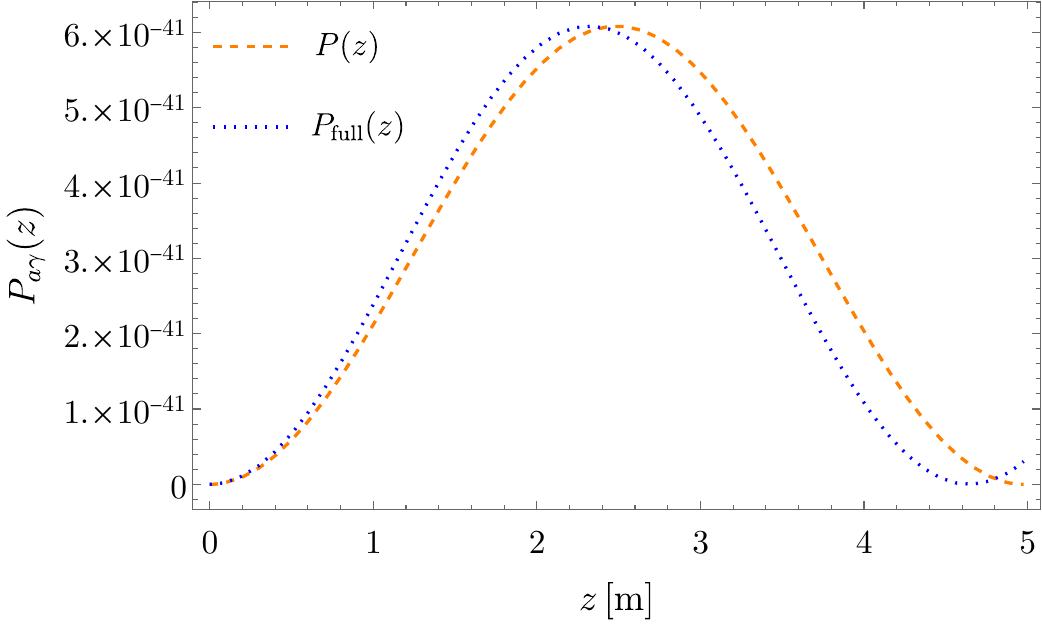}
        \label{fig:comp_nr}

        \includegraphics[width=0.9\columnwidth]{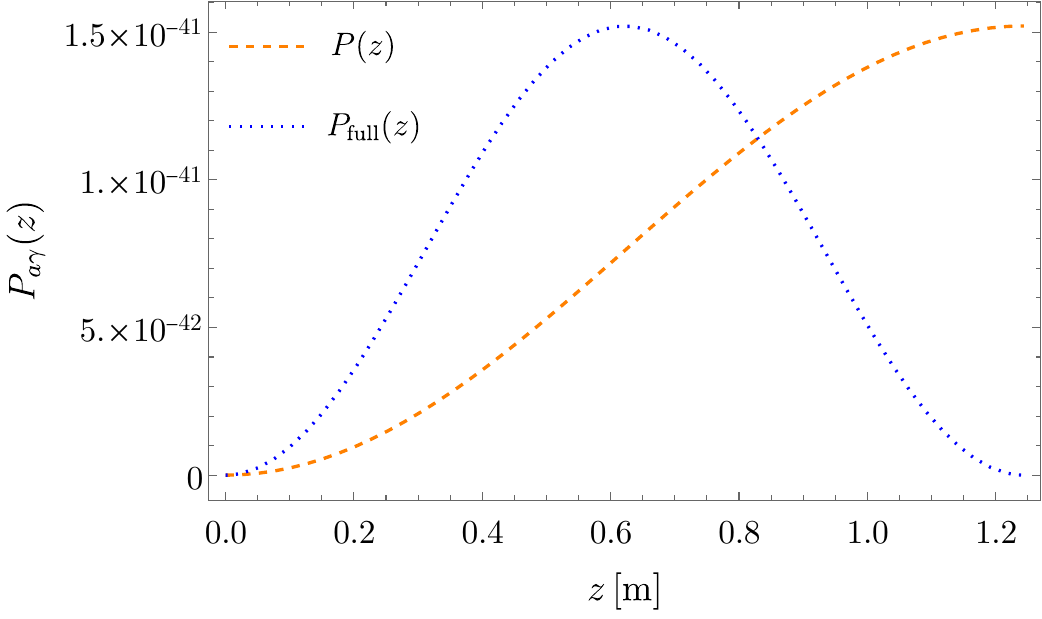}
        \label{fig:comp_1ma}
    \caption{ Comparison between Eq.~\eqref{eq:probrel}, shown in dashed orange lines, and Eq.~\eqref{eq:finalPfull}, in dotted blue lines. We used $B = 10^{-6} \text{G}$, $g_{a\gamma} = 10^{-10}\text{GeV}^{-1}$ and  $m_a = 10^{-6}\,\text{eV}$. The top panel shows $\omega = 100\,m_a$, the middle panel $\omega = 2\,m_a$, and the bottom one $\omega = 1.001m_a$. We see that in the relativistic limit (top panel), the agreement between Eqs.~\eqref{eq:probrel} and \eqref{eq:finalPfull} is complete, while small deviations start to arise in the mildly relativistic regime (middle panel). In the completely nonrelativistic limit (bottom panel), the predictions of the two formulae are completely different. We notice that, even in the nonrelativistic limit, Eq.~\eqref{eq:probrel} correctly predicts the amplitude of the oscillations, as it can be understood by comparing it to Eq.~\eqref{eq:finalPfull}.  The size of the $x$-axis is given by 2$\pi$($\Delta_\text{osc}^\text{full})^{-1}$, where $\Delta_\text{osc}^\text{full}$ is given by Eq.~\eqref{eq:doscfull}. In this setting $Q_a \gg Q_\parallel, Q_M$, so the relativistic limit is correctly identified by $\omega\gg m_a$.}
   
    \label{fig:comparison}
\end{figure}

\begin{figure}[t!]
        \label{fig:osc_rel}

        \includegraphics[width=0.9\columnwidth]{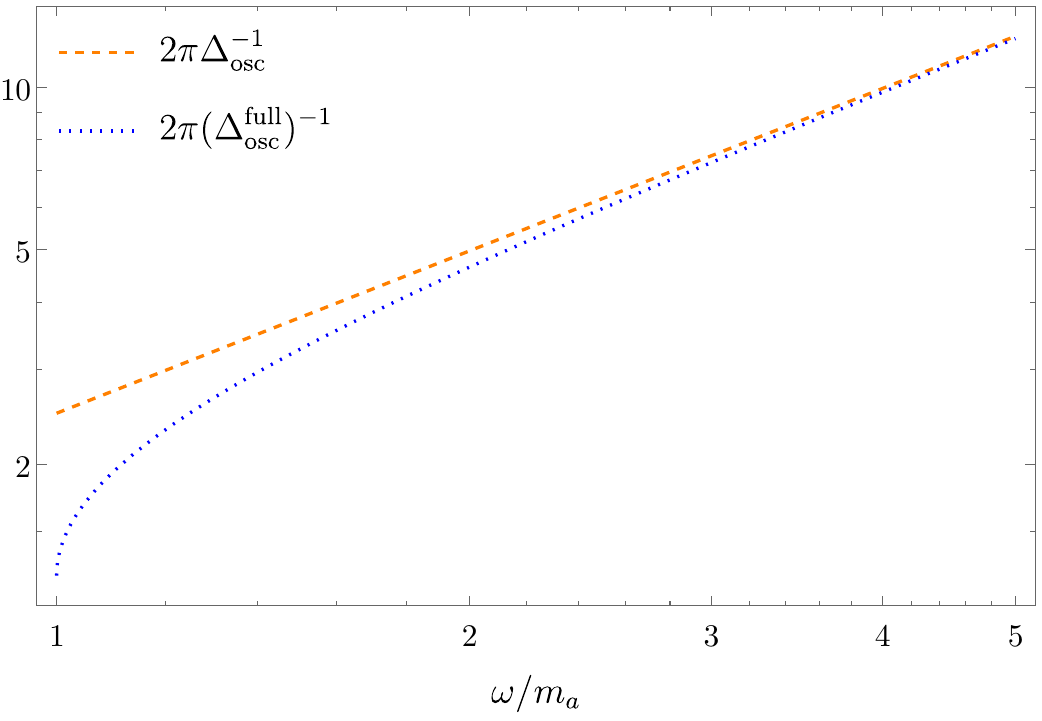}
        \label{fig:osc_nr}
    \caption{Comparison between the oscillation lengths defined by Eqs.~\eqref{eq:doscrel} and \eqref{eq:doscfull}, plotted as a function of frequency $\omega$ (normalized by the axion mass $m_a$) for the same set of parameters as Fig.~\ref{fig:comparison}. As discussed in the main text, they are indistinguishable in the relativistic limit. 
    However, in the nonrelativistic limit, the two expressions differ. The deviation is shown with logarithmic axes for a better visualization, and is reflected also in the middle and bottom panels of Fig.~\ref{fig:comparison}.}
    \label{fig:osc_lengths}
\end{figure}

In Sec.~\ref{sec:basics}, we critically re-examined the derivation of the axion-photon conversion probability in the relativistic limit. However, we recognize that we do not need to specify to the relativistic limit to solve the equations. Indeed, they can be solved in full generality by plugging the ansatz in Eq.~\eqref{eq:pw} into Eq.~\eqref{eq:simpleom}. In this case, we get
\begin{equation}
     \begin{pmatrix}
        k'_\parallel\\
        k'_a
    \end{pmatrix} = \omega
    \begin{pmatrix}
        \sqrt{1 + Q'_\parallel/\omega^2}\\
        \sqrt{1 + Q'_a/\omega^2}
    \end{pmatrix}\equiv\begin{pmatrix}
        \Delta'_{\parallel,f}\\
        \Delta'_{a,f}
    \end{pmatrix},
    \label{eq:ksfull}
\end{equation}
which reduces to Eq.~\eqref{eq:ks} in the relativistic limit, after eliminating the term proportional to the identity matrix, which amounts to an irrelevant phase.\footnote{You can also see it from a different angle. Plugging Eq.~\eqref{eq:pw} in Eq.~\eqref{eq:t1before}, i.e. before redefining the field $\Psi'$, leads to an equation analogous to Eq.~\eqref{eq:ks}, which is exactly the relativistic limit of Eq.~\eqref{eq:ksfull}. Anyway, common phases are irrelevant as for the determination of conversion probabilities.}
Incidentally, we also clearly understand that the relativistic approximation carried out in the previous section requires $\omega^2 \gg |Q'_\parallel|, |Q_a'|$, and not only $\omega \gg m_a$. It is really the statement $\omega \gg M$, where $M$ is a placeholder for $|Q'_\parallel|^{1/2}, |Q_a'|^{1/2}$. Thus, we discover that Eq.~\eqref{eq:ksfull} reduces to Eq.~\eqref{eq:ks} in the relativistic limit and, in that limit, we get exactly the result in Eq.~\eqref{eq:probrel}.\\
In the general case, instead, we have
\begin{align}
    \begin{pmatrix}
        A_\parallel(z)\\
        a(z)
    \end{pmatrix} &= \mathcal{O}^{-1}   
    \begin{pmatrix}
        e^{-i\Delta'_{\parallel,f} z}&0\\
        0&e^{-i\Delta'_{a,f} z}\\
    \end{pmatrix}
    \mathcal{O}
        \begin{pmatrix}
        A_\parallel\\
        a
    \end{pmatrix} \nonumber\\
    &\equiv M_\text{full}(z)
    \begin{pmatrix}
        A_\parallel\\
        a
    \end{pmatrix},
    \label{eq:evol_full}
\end{align}
in analogy with Eq.~\eqref{eq:evol}. In this setting, the conversion probability reads
\begin{align}
    P_\text{full}(z) =& \lvert M_\text{full,12}\rvert^2\nonumber\\
    =&\left| \cos\theta\sin\theta \left[e^{-i\Delta'_{\parallel,f} z}-e^{-i\Delta'_{a,f} z}\right]\right|^2\,.
    \label{eq:Pfull}
\end{align}
Now, we can rewrite the result, obtaining
\begin{align}
     P_\text{full}(z) = &\sin^22\theta\sin^2\left[\frac{1}{2}\left(\Delta'_{\parallel,f} - \Delta'_{a,f} \right)z\right]\nonumber\\
     =&\frac{4Q_{a\gamma}^2}{\Delta Q^2}\sin^2\left[\frac{1}{2}\left(\Delta'_{\parallel,f} - \Delta'_{a,f} \right)z\right],
     \label{eq:finalPfull}
\end{align}
where, analogously to Eq.~\eqref{eq:doscrel}, we can define an oscillation length as 
\begin{equation}
    \Delta_\text{osc}^\text{full} = \Delta'_{\parallel,f} - \Delta'_{a,f}.
    \label{eq:doscfull}
\end{equation}
Equations~\eqref{eq:finalPfull} and \eqref{eq:doscfull} are the general formulae for the axion-photon conversion, valid in every energy regime (under the assumption of constant magnetic field and homogeneous plasma). They reduce to Eqs.~\eqref{eq:probrel} and \eqref{eq:doscrel} in the relativistic case and they are also valid in the nonrelativistic scenario.\\
In particular, it is easy to see, employing the definitions in Eqs.~\eqref{eq:dq} and \eqref{eq:di}, that the relativistic formula in Eq.~\eqref{eq:probrel} exactly reproduces the amplitude of the oscillations predicted from Eq.~\eqref{eq:finalPfull}. However, the oscillation length will in general be different. For instance, if we consider $Q_a \gg Q_\parallel, Q_M$ and $\omega \sim m_a$, Eqs.~\eqref{eq:probrel} and \eqref{eq:finalPfull} reduce to 
\begin{align}
    &P_\text{rel}(z)  \sim \frac{4Q^2_{a\gamma}}{Q_a^2}\sin^2\left(\frac{m_a}{4}z\right),\\
    &P_\text{full}(z)  \sim \frac{4Q^2_{a\gamma}}{Q_a^2}\sin^2\left(\frac{m_a}{2}z\right),
\end{align}
which complies with numerical findings. Precisely, in Fig.~\ref{fig:comparison} we compare the results of the exact formula in Eq.~\eqref{eq:finalPfull} (blue dotted line) with its relativistic limit in Eq.~\eqref{eq:probrel} (orange dashed line). In the upper panel the two formulae are in agreement because the axion energy is $\omega=100\,m_{a}$, safely within the relativistic limit. The middle panel shows the mildly relativistic case $\omega=2\,m_{a}$, where the two probabilities have matching amplitudes, but are out of phase because of the shorter oscillation length of the exact result.  In the lower panel we show the nonrelativistic case $\omega=1.001\,m_{a}$, where the discrepancy between the two cases increases even more.
The behavior of the oscillation length in the two formulations, as a function of the energy, is shown in Fig.~\ref{fig:osc_lengths}. The exact formulation (blue dotted line) always predicts a shorter oscillation length compared to the relativistic approximation (orange dashed line), which is manifestly inappropriate for $\omega\lesssim 2\,m_{a}$.

\section{Perturbative treatment of inhomogeneity}
\label{sec:pert}
If the plasma is inhomogeneous, or if the magnetic field is not constant and has a characteristic scale of variation, then the solution found in Eq.~\eqref{eq:finalPfull} is not valid anymore. However, we can perturbatively study the regime in which the inhomogeneity is very small. In other words, we can restrict to a situation in which the typical length scale of the perturbation is much larger than the one characterizing the axion-photon conversion [e.g. see Eq.~\eqref{eq:doscfull}].
In this case, Eq.~\eqref{eq:simpleom} reduces to 
\begin{align}
    &\left[\omega^2 + \partial_z^2 + 
    \mathcal{O}^{-1}\mathcal{M}^2_\text{diag}\mathcal{O}\right]\Psi = 0,\\
    &\left[\omega^2 + \mathcal{O}\partial_z^2+ 
    \mathcal{M}^2_\text{diag}\mathcal{O}\right]\mathcal{O}^{-1}\Psi' = 0\label{eq:Mdiag_inh},
\end{align}
where the second line is obtained by multiplying the first by $\mathcal{O}$ and the partial derivative is acting on the product  $\mathcal{O}^{-1}\Psi'$. In this setup, both the diagonal mass matrix $\mathcal{M}_\text{diag}^2$ and the mixing angle $\theta$ are inherently dependent on $z$. At each value of $z$, however, they can be defined exactly as in Eqs.~\eqref{eq:thetamat} and \eqref{eq:m2diagonal}.
To proceed, we define
\begin{equation}
     \partial_z\mathcal{O}^{-1} = \partial_z\begin{pmatrix}
        \cos\theta&-\sin\theta\\
        \sin\theta&\cos\theta\\
    \end{pmatrix}\equiv -\theta_z
    \begin{pmatrix}
        \sin\theta&\cos\theta\\
        -\cos\theta&\sin\theta\\
    \end{pmatrix},
\end{equation}
where $\theta_z \equiv \partial_z\theta$ takes into account the inhomogeneity of the plasma (or of the magnetic field). The smallness of the perturbation can be interpreted as $\lvert\theta_z\rvert < \Delta_\text{osc}^\text{full}$, where $\Delta_\text{osc}^\text{full}$ is the axion-photon conversion length scale defined in Eq.~\eqref{eq:doscfull}.
This naturally leads us to interpret $\theta_z$ as an expansion parameter,\footnote{Actually, $\theta_z$ has the dimensions of an energy. Therefore, the proper expansion parameter would be that represented by the dimensionless combination $\theta_z/\Delta_\text{osc}^\text{full}$. We will understand this subtlety in most of the analysis.} so that we can retain only the first-order term in the expansion and fix $\theta_z$ to $\theta_z(z=0)$. On this line, we can derive
\begin{align}
    &\mathcal{O}\partial_z\mathcal{O}^{-1} =  -\theta_z
    \begin{pmatrix}
        0&1\\
        -1&0\\
    \end{pmatrix},\label{eq:dzO}\\
    &\mathcal{O}\partial_z^2\mathcal{O}^{-1} = -\theta_{zz}
    \begin{pmatrix}
        0&1\\
        -1&0\\
    \end{pmatrix} -\theta_z^2
        \begin{pmatrix}
        1&0\\
        0&1\\
    \end{pmatrix}\sim 0,\label{eq:d2zO}     
\end{align}
where $\theta_{zz} \equiv \partial_z^2 \theta$ and $\mathcal{O}\partial_z^2\mathcal{O}^{-1}\sim 0$, as it is higher order in the $\theta_z$ expansion.
In total, Eq.~\eqref{eq:Mdiag_inh} reads:
\begin{equation}
\begin{split}
    &\left[\omega^2 + \partial_z^2+ 
    \mathcal{M}^2_\text{diag}\right]\Psi' + 2\mathcal{O}\partial_z \mathcal{O}^{-1}\partial_z \Psi' + \\
    &+\mathcal{O}\partial_z^2\mathcal{O}^{-1}\Psi'= 0
\end{split}
\end{equation}
or, using Eq.~\eqref{eq:d2zO},
\begin{equation}
\begin{split}
    \left[\omega^2 + \partial_z^2+ 
    \mathcal{M}^2_\text{diag}\right]\Psi' + 2\mathcal{O}\partial_z \mathcal{O}^{-1}\partial_z \Psi' = 0.\label{eq:syst_inh}
\end{split}
\end{equation}
This equation can be compared with Eq.~\eqref{eq:eomdiag}, valid in the case of $\theta_z = 0$. In the inhomogeneous setting, it is immediate to see that the components $A_\parallel'$ and $a'$, encapsulated in the vector $\Psi'$, will no longer be propagation eigenstates. In contrast to the constant-$\theta$ case, they will mix, with a strength that will, in general, be proportional to $\theta_z$.
Writing the system in Eq.~\eqref{eq:syst_inh} explicitly, we have
\begin{equation}
    \begin{split}
        &\left[\omega^2 + \partial_z^2 + Q_\parallel' \right]A_\parallel' - 2\theta_z \partial_z a' = 0,\\
        &\left[\omega^2 + \partial_z^2 + Q_a'\right]a' + 2\theta_z \partial_z A' = 0,
    \end{split}
    \label{eq:inhom}
\end{equation}
where we have understood the $z$ dependence of $Q_\parallel', Q_a'$ for clarity.
We can solve this set of equations perturbatively. In particular, we can take $Q_\parallel'(z) \sim Q_\parallel'(z = 0)\equiv Q_{\parallel,0}'$ and $Q_a'(z) \sim Q_a'(z=0)\equiv Q_{a, 0}'$. At this point, we firstly solve the homogeneous system
\begin{equation}
    \begin{split}
        &\left[\omega^2 + \partial_z^2 + Q_{\parallel,0}'\right]A_{\parallel,h}' = 0,\\
        &\left[\omega^2 + \partial_z^2 + Q_{a, 0}'\right]a_h'  = 0,
    \end{split}
    \label{eq:hom}
\end{equation}
which, using Eq.~\eqref{eq:ksfull}, gives
\begin{equation}
\begin{pmatrix}
        A'_{\parallel,h}(z)\\
        a'_h(z)
\end{pmatrix} = \begin{pmatrix}
    \bar{A}'_{\parallel,h}\, e^{-i\Delta_{\parallel,f}' z}\\
    \bar{a}'_h\, e^{-i\Delta_{a,f}' z}
\end{pmatrix},
\label{eq:pwfull}
\end{equation}
with $ \bar{A}'_{\parallel,h}, \bar{a}'_h$ constant.
To be precise, we should have inserted a subscript ``0" in $\Delta'_{a,f}$ and $\Delta'_{\parallel,f}$ to signal that they are computed at the initial point $z=0$ with $Q_{a, 0}', Q_{\parallel, 0}'$. However, we omit this to avoid cluttering the notation.
Then we plug the solution of Eq.~\eqref{eq:hom} in the $\theta_z$-dependent terms in Eq.~\eqref{eq:inhom}. With this approximation, the system in Eq.~\eqref{eq:inhom} reduces to a well-known class of second-order differential equations:
\begin{equation}
    y''(z) + c_1 y'(z) + c_2 y(z) = g(z)
    \label{eq:diff_class}
\end{equation}
with $c_1, c_2 \in \mathbb{R}$ and $g(z) = Q(z)e^{\lambda z}$. The full solution of the differential equation in Eq.~\eqref{eq:diff_class} can be expressed as the sum of the homogeneous and particular solutions, which, for Eq.~\eqref{eq:inhom}, yields
\begin{equation}
\begin{pmatrix}
        A'_\parallel(z)\\
        a'(z)
\end{pmatrix} = \begin{pmatrix}
    \bar{A}'_{\parallel,h}\, e^{-i\Delta_{\parallel,f}' z} + \bar{A}'_{\parallel,p} e^{-i\Delta_{a,f}' z}\\
    \bar{a}_h'\, e^{-i\Delta_{a,f}' z} + \bar{a}_p'\,e^{-i\Delta_{\parallel,f}' z} 
\end{pmatrix},
\label{eq:fullsol}
\end{equation}
with $\bar{A}'_{\parallel,h}, \bar{A}'_{\parallel,p}, \bar{a}_h'$ and $ \bar{a}_p'$ constant. By substitution of Eq.~\eqref{eq:fullsol} into Eq.~\eqref{eq:inhom}, we can find the coefficients $\bar{A}'_{\parallel,p}$ and $\bar{a}_p'$, which read
\begin{equation}
    \begin{split}
    &\bar{A}'_{\parallel,p} = 2i\theta_z \frac{\Delta_{a,f}'}{Q_{a,0}' - Q_{\parallel,0}'} \bar{a}_h' \equiv M \bar{a}_h',\\ 
    &\bar{a}_p' = 2i\theta_z \frac{\Delta_{\parallel,f}'}{Q_{a,0}' -Q_{\parallel,0}'}\bar{A}'_{\parallel,h}\equiv N \bar{A}'_{\parallel,h},
    \end{split}
    \label{eq:defs}
\end{equation}
where we define some order-$\theta_z$ auxiliary quantities $M, N$ for later convenience.
Notice that the particular solution is of order $\theta_z$, which justifies a-posteriori the approach followed before. Finally, we can rewrite Eq.~\eqref{eq:fullsol} as 
\begin{equation}
\begin{split}
    \begin{pmatrix}
        A'_\parallel(z)\\
        a'(z)
\end{pmatrix} = &\begin{pmatrix}
     e^{-i\Delta_{\parallel,f}' z} &    Me^{-i\Delta_{a,f}' z}\\
    Ne^{-i\Delta_{\parallel,f}' z} & e^{-i\Delta_{a,f}' z} 
\end{pmatrix} \begin{pmatrix}
        \bar{A}'_{\parallel,h}\\
        \bar{a}'_h
\end{pmatrix}\\
\equiv &\,S(z)\begin{pmatrix}
        \bar{A}'_{\parallel,h}\\
        \bar{a}'_h
\end{pmatrix}.
\end{split}
\label{eq:halfway}
\end{equation}
As $\bar{A}'_{\parallel,h}$ and $\bar{a}'_h$ are $z$-independent quantities, Eq.~\eqref{eq:halfway} can be seen as expressing the fields $A_\parallel'$ and $ a'$ at the point $z$ as a function of $z=0$ quantities. However, initial conditions on photons (axions) are given on the full $A'_\parallel$ ($a'$), not on $\bar{A}'_{\parallel,h}$ ($\bar{a}'_h$). Therefore, we should incorporate this requirement in Eq.~\eqref{eq:halfway}. In particular, evaluating Eq.~\eqref{eq:fullsol} at $z = 0$ and using Eq.~\eqref{eq:defs} yields
\begin{equation}
\begin{pmatrix}
        A'_\parallel(0)\\
        a'(0)
\end{pmatrix} = \begin{pmatrix}
    \bar{A}'_{\parallel,h} +  M \bar{a}_h'\\
    \bar{a}_h'+  N \bar{A}'_{\parallel,h}
\end{pmatrix},
\label{eq:fullsol0}
\end{equation}
which can be inverted to give
\begin{equation}
\begin{split}
\begin{pmatrix}
    \bar{A}'_{\parallel,h}\\
        \bar{a}'_h\end{pmatrix} = &\begin{pmatrix}
    A'_{\parallel}(0) -   M a'(0)\\
    a'(0) - N A'_{\parallel}(0)
\end{pmatrix}\\
\equiv&\, L\begin{pmatrix}
    A'_{\parallel}(0)\\
        a'(0)\end{pmatrix},
\end{split}
\end{equation}
valid up to first order in $\theta_z$.
Therefore, the solution is
\begin{equation}
     \begin{pmatrix}
        A'_\parallel(z)\\
        a'(z)
    \end{pmatrix} = \,S(z)\, L     \begin{pmatrix}
        A'_\parallel(0)\\
        a'(0)
    \end{pmatrix} 
\end{equation}
and rotating back to the initial fields we finally have
\begin{equation}
\begin{split}
     \begin{pmatrix}
        A_\parallel(z)\\
        a(z)
    \end{pmatrix} &= \mathcal{O}^{-1}(z)\,S(z)\, L\, \mathcal{O}(0)     \begin{pmatrix}
        A_\parallel(0)\\
        a(0)
    \end{pmatrix}\equiv\\
    &\equiv M_\text{pert}(z) \begin{pmatrix}
        A_\parallel(0)\\
        a(0)
    \end{pmatrix}.
    \label{eq:assafa} 
\end{split}
\end{equation}
The evolution matrix $M_\text{pert}$ describes the dynamics of the axion-photon mixing when the plasma or magnetic field inhomogeneities can be treated as a small perturbation. Analogously to what is done in Eqs.~\eqref{eq:probrel} and \eqref{eq:finalPfull}, we can obtain an analytic form that well approximates the conversion probability. Indeed, at leading order, the conversion probability reads
\begin{equation}
\begin{split}
P_\text{pert}(z)& = \cos(\theta_0)^2 \sin(\theta)^2 + \cos(\theta)^2 \sin(\theta_0)^2 \\
&\quad-
\frac{1}{2} \cos\left[\left(\Delta_{a, f}' -\Delta_{\parallel, f}' \right) z\right] \sin(2 \theta) \sin(2 \theta_0)\,,
\label{eq:Ppert}
\end{split}
\end{equation}
where $\theta$ is evaluated at $z$, while $\theta_0$ is the initial value of the mixing angle, i.e., $\theta_0 \equiv \theta(z = 0)$. 

\begin{figure}[t!]
    \centering
    \vspace{4ex}%
    \includegraphics[width=0.46\textwidth]{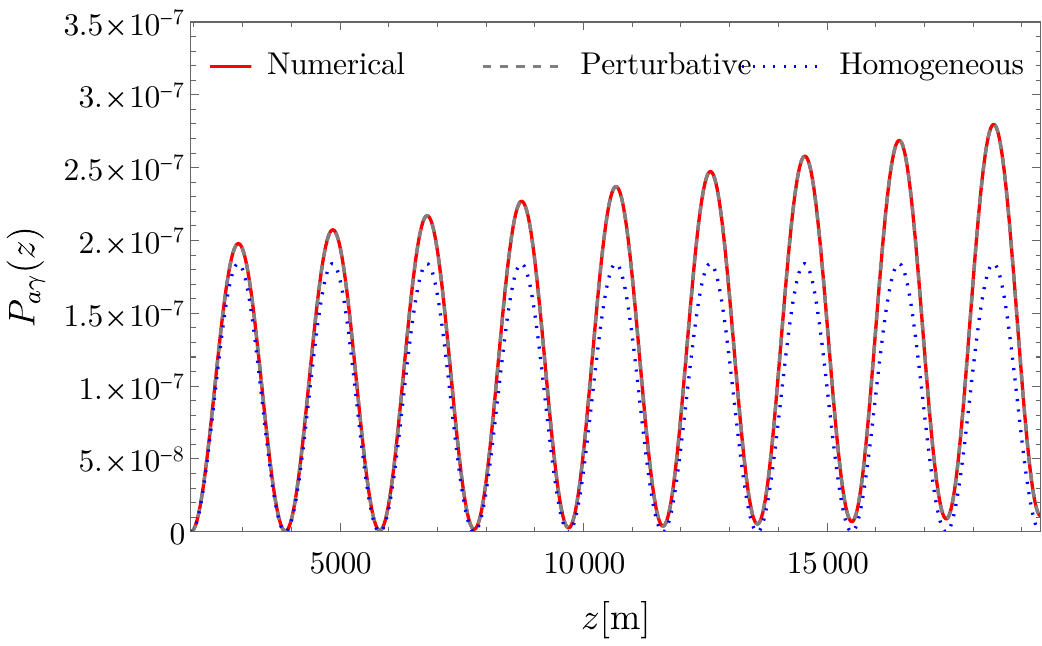}
    \includegraphics[width=0.46\textwidth]{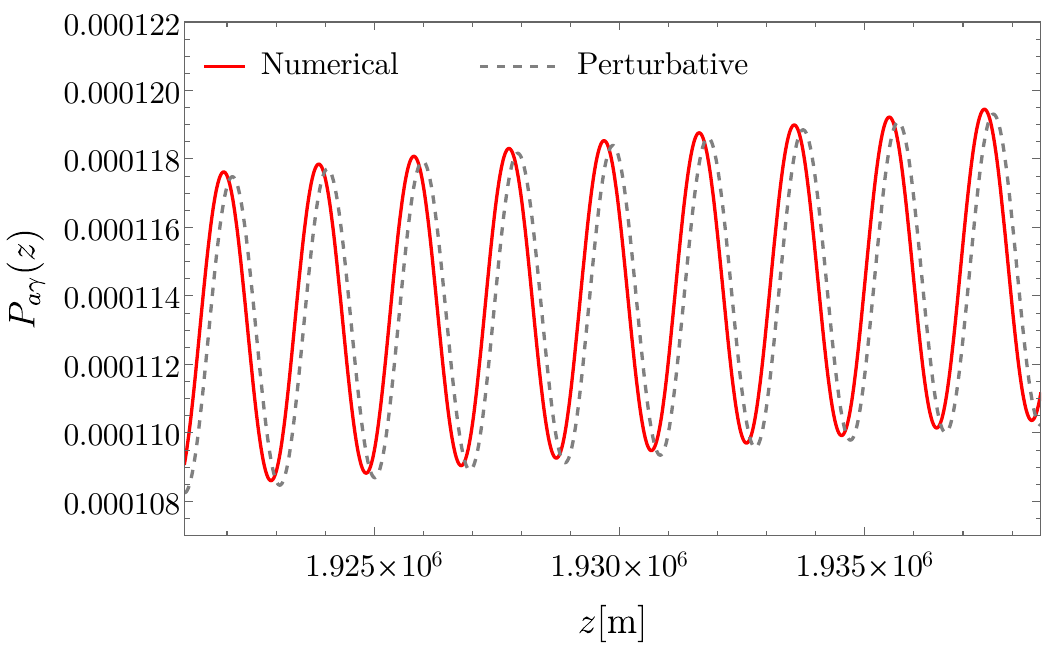}
    \caption{Plot of the axion-photon conversion probability. We consider an axion with mass $m_a = 10^{-9}~\text{eV}$, coupling $g_{a\gamma} = 10^{-14}\, \text{GeV}^{-1}$ and energy $\omega = 1.1~m_a$, traveling in a magnetic field  $B(z) \sim B_0 (1 +  10^{-2}\Delta_\text{osc}^\text{full} z )$, with $B_0 = 10^{12}~\text{G}$. The solid red line shows the numerical result, based on the solution to Eq.~\eqref{eq:simpleom}. The dotted blue line shows the homogeneous analytical formula in Eq.~\eqref{eq:finalPfull}, valid when both the plasma and the magnetic field are homogeneous. Finally, the dashed gray line represents the perturbative approximation determined in Eq.~\eqref{eq:Ppert}, valid when the plasma or the magnetic field feature small inhomogeneities.
    The top panel shows the comparison for values of $z$ well within the approximation validity range (see main text).
    The bottom panel, instead, shows the comparison for values around $z \sim 10^6~\text{m}$, where the perturbative expansion is expected to fail. Here, we do not plot the homogeneous solution, which oscillates with the same amplitude and frequency as shown by the top panel.}
        \label{fig:pcomp0}
\end{figure}

Using simple trigonometric identities, it is easy to see that Eq.~\eqref{eq:Ppert} reduces to Eq.~\eqref{eq:finalPfull} in the homogeneous case ($\theta\rightarrow \theta_0$).
In Fig.~\ref{fig:pcomp0}, we plot the conversion probability obtained from the numerical solution (solid red line) of the fully $z$-dependent Eq.~\eqref{eq:simpleom}, comparing it with the homogeneous [Eq.~\eqref{eq:finalPfull}, in dotted blue] and the perturbative [Eq.~\eqref{eq:Ppert}, in dashed gray] solutions, obtained, respectively, when the magnetic field is homogenous or features small inhomogeneities.
To produce Fig.~\ref{fig:pcomp0}, we consider a magnetic field varying on a length scale much smaller than $\Delta_\text{osc}^\text{full}$ defined in Eq.~\eqref{eq:doscfull}. In particular, we select an axion with mass $m_a = 10^{-9}~\text{eV}$, coupling $g_{a\gamma} = 10^{-14}\,\text{GeV}^{-1}$ and energy $\omega = 1.1~m_a$, traveling in a magnetic field $B(z) \sim B_0 (1 +  10^{-2}\Delta_\text{osc}^\text{full} z )$, with $B_0 = 10^{12}~\text{G}$. In principle, one could repeat the analysis for plasma frequency inhomogeneities as well.

In particular, Fig.~\ref{fig:pcomp0} shows that the perturbative approach provides a first attempt to track the variation of the conversion amplitude, induced by a varying magnetic field.
Indeed, the change in the amplitude is mainly driven by the modification in $\theta$, which grows as $B(z)$ becomes greater for increasing values of $z$.

As the perturbative solution requires constant $Q'_{a,0}$ and $ Q'_{\parallel,0}$ values, the range of validity of the approximation in Eq.\eqref{eq:fullsol} can be found by demanding that, roughly,  
\begin{equation}
\begin{split}
    &\text{min}\left[\left(k'_\parallel(z) - k'_{\parallel,0}\right) ,\left(k'_a(z) - k'_{a,0}\right)\right] \cdot z\ll 2\pi,\\
    &\theta_{zz} \ll \theta_z \Delta^{\text{full}}_\text{osc}\,.
\end{split}
    \label{eq:val}
\end{equation}
In other words, we demand that the phase difference between the perturbative and the numerical solutions does not exceed $2\pi$. The perturbative approximation is broken at the distance $z$ that saturates this condition.

Moreover, as we neglect higher-order terms in $\theta_z$, we should also ensure that the second condition in Eq.~\eqref{eq:val} is satisfied.
For the example in Fig.~\ref{fig:pcomp0}, we find $z \lesssim 10^{6}~\text{m}$, which is confirmed numerically.

\section{Resonant conversion}
\label{sec:LZ}

In this section, we discuss resonant axion-photon conversions happening when $Q_{a}=Q_{\parallel}$. We consider the case $B_{0}\ll B_{\rm crit}$, leading to the resonant condition being satisfied when the axion mass is equal to the plasma frequency. Resonant conversions are sensitive on the local properties of the plasma frequency and magnetic field at the resonance point. This feature is in contrast with nonresonant conversions, which depend on the properties of the plasma along the axion-photon beam path.
In particular, resonances are of primary importance in the case of DM axions converting into radio waves in a neutron star magnetosphere~\cite{Pshirkov:2007st} and for conversions happening in the primordial magnetic field, which cause distortions in the cosmic microwave background blackbody spectrum~\cite{Mirizzi:2009nq,Mukherjee:2018oeb}.
We can have an analytical understanding of resonant conversions focusing on a constant magnetic field and a linearly varying $\omega_{\rm pl}^{2}$. In this setting, we can express the conversion probability in terms of a generalized Landau-Zener (LZ) formula~\cite{Carenza:2023nck} (see also Refs.~\cite{Landau:1932vnv,Majorana:1932ga,Stueckelberg:1932,Zener:1932ws,Landau:1991wop}). 

\begin{figure}[t!]
    \centering
    \vspace{4ex}%
    
    \includegraphics[width=0.46\textwidth]{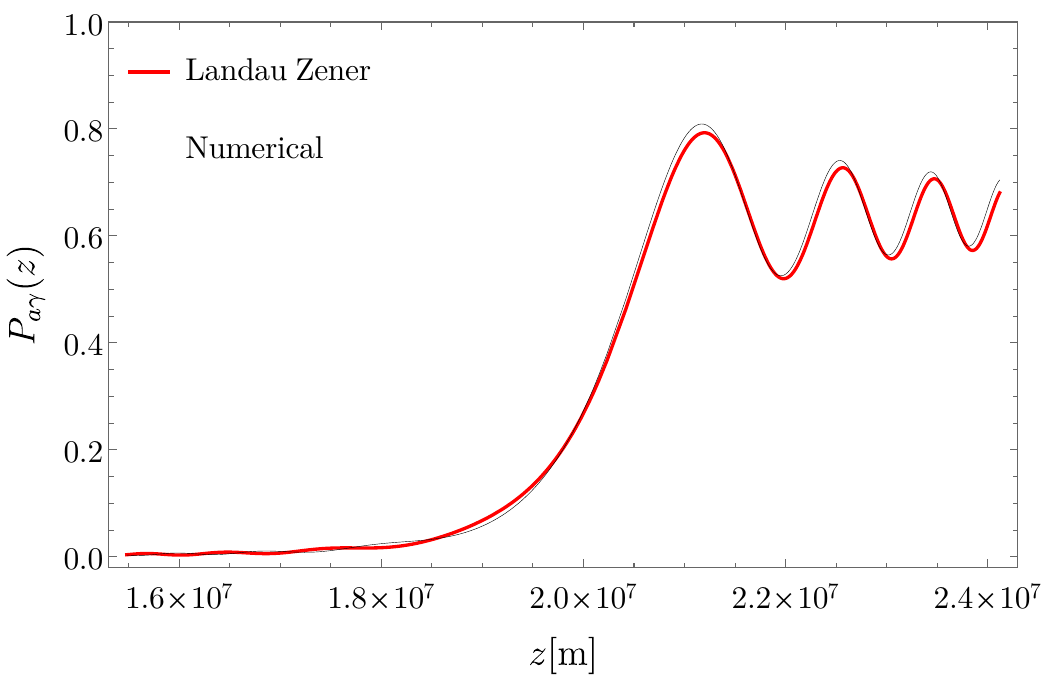}
    \includegraphics[width=0.46\textwidth]{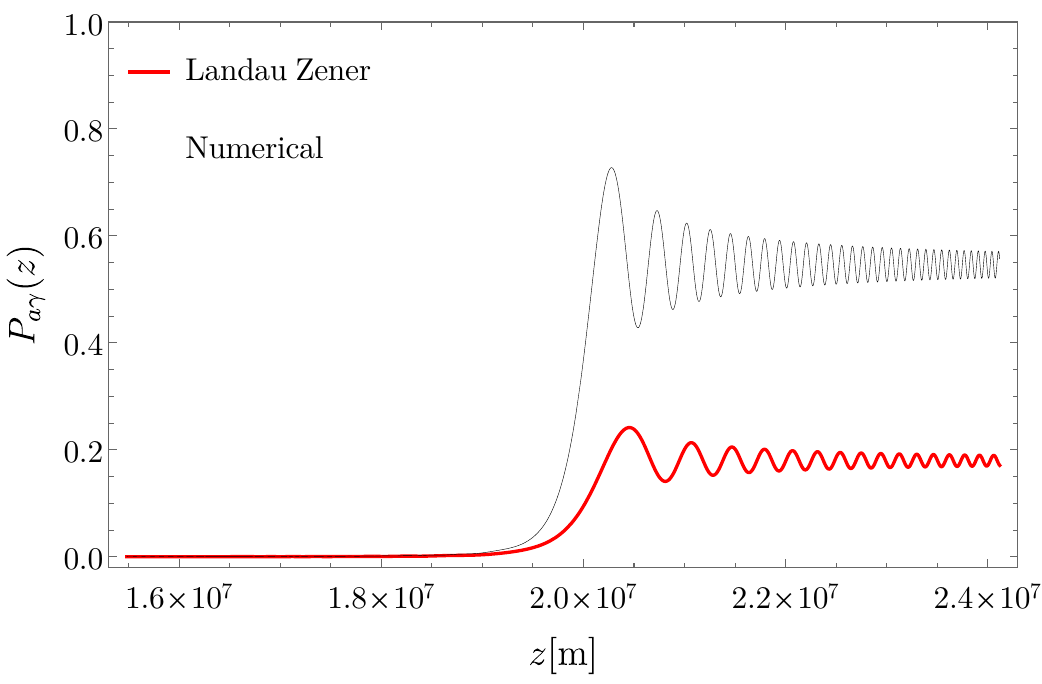}
    \caption{Plot of the axion-photon conversion probability as a function of the traveled distance $z$. We consider an axion with mass $m_a = 10^{-10}\, \text{eV}$, coupling $g_{a\gamma} = 10^{-14}\,\text{GeV}^{-1}$ traveling in a constant magnetic field $B_0 = 2\times 10^{12}\,\text{G}$, and varying plasma frequency. When $Q_a = Q_\text{par}$, which happens at $z = z_0$, level crossing takes place. In the top panel, we show the numerical solution (black line) compared to the analytic approximation (red line), computed in Ref.\cite{Carenza:2023nck}, for a relativistic axion with energy $\omega = 6\,m_a$. The bottom panel displays the comparison for an almost nonrelativistic axion with energy $\omega = 1.2\, m_a $ instead. See the main text for more details. }
    \label{fig:LZ1}
\end{figure}

\begin{figure}[t!]
    \includegraphics[width=0.46\textwidth]{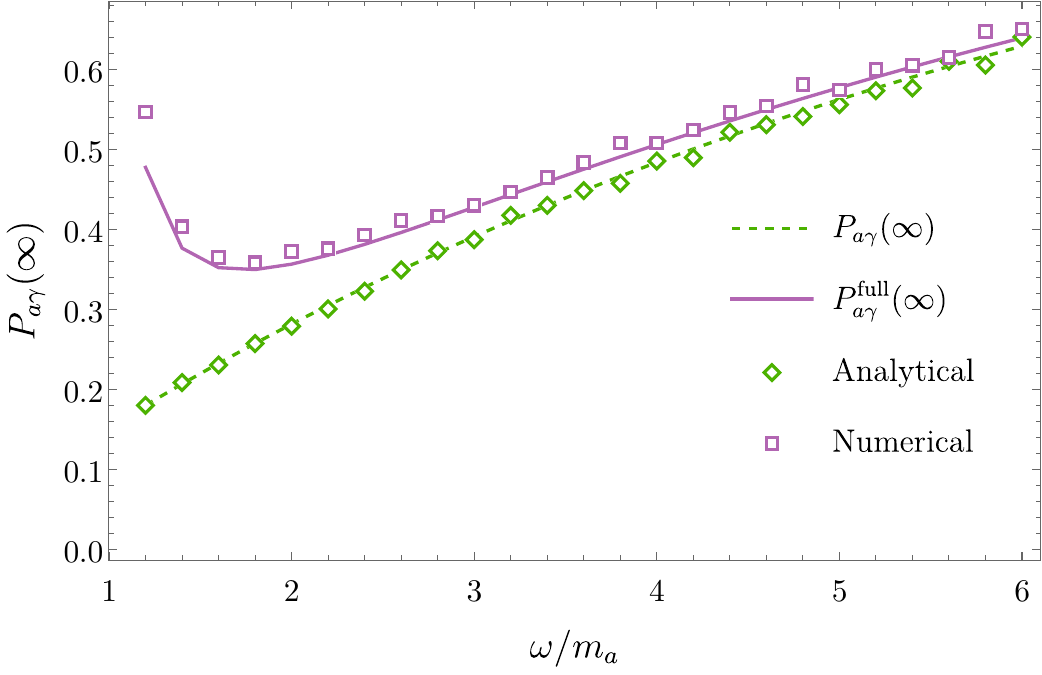}
    \caption{Plot of the asymptotic axion-photon conversion probability as a function of the axion energy. We consider an axion with mass $m_a = 10^{-10}\, \text{eV}$, coupling $g_{a\gamma} = 10^{-14}\,\text{GeV}^{-1}$ traveling in a constant magnetic field $B_0 = 2\times 10^{12}\,\text{G}$, and varying plasma frequency. 
    The numerical solution to the system in Eq.~\eqref{eq:simpleom} is shown using violet squares. 
    The analytic computation in Ref.~\cite{Carenza:2023nck} (green rhombi), based on the Schr\"odinger-like picture described in Sec.~\ref{sec:basics}, correctly captures the numerical asymptotic conversion probability for relativistic axions, while deviates in the mildly to non relativistic regime, leading to a wrong prediction. We also show the LZ formula in Eq.~\eqref{eq:estimate}, labeling it as $P_{a\gamma}(\infty)$(dashed green line) and $P_{a\gamma}^\text{full}(\infty)$ (solid violet line), computed using the oscillation lengths $\Delta_\text{osc}$ and $\Delta_\text{osc}^\text{full}$, respectively.}
        \label{fig:pcomp}
\end{figure}

In the following, we consider this scenario and compare the exact numerical solution of Eq.~\eqref{eq:simpleom} with the analytical results in the relativistic axion case, as given by Eq.~(3.17) in Ref.~\cite{Carenza:2023nck}. Our findings are shown in Fig.~\ref{fig:LZ1}, for a plasma frequency expressed as
\begin{equation}
    Q_{\rm par}=-m_{a}^{2}\left(\frac{z}{z_{0}}\right)\,,
\end{equation}
with a resonance, or level crossing, at $z=z_{0}\sim 2\times10^{7}$~m. To produce Fig.~\ref{fig:LZ1}, we consider an axion with mass $m_a = 10^{-10}\, \text{eV}$, coupling $g_{a\gamma} = 10^{-14}\,\text{GeV}^{-1}$ and traveling in a constant magnetic field $B_0 = 2\times 10^{12}\,\text{G}$. The two panels display the conversion probability for a relativistic axion, with energy $\omega=6\,m_{a}$ (upper panel), and an almost nonrelativistic one, with energy $\omega=1.2\,m_{a}$ (lower panel). The exact solution is given by the thin black line, while the red line corresponds to the analytical relativistic approximation presented in Eq.~(3.17) in Ref.~\cite{Carenza:2023nck}. In the first case, the two solutions show a good agreement, as expected; in the mildly relativistic case, instead, the relativistic approximation significantly underestimates the numerical result. 
We can highlight this mismatch focusing on the behavior of the conversion probability at a large distance from the resonance point, as a function of the axion energy, as shown in Fig.~\ref{fig:pcomp}. Here, the numerical exact solution (violet squares) always overestimates the analytical relativistic formula (green rhombi), up to an order-one disagreement at the lowest plotted energy. These results suggest that one should be careful in applying the relativistic axion formulation to mildly to nonrelativistic axions, because there are important disagreements. 
We can qualitatively understand this behavior as follows.
Firstly, we notice that, in the case of mildly relativistic axions, the oscillation length is smaller than what would have been predicted by a naive application of the relativistic approximation (see Fig.~\ref{fig:osc_lengths}). When applying the LZ formula, the conversion probability at a large distance from the resonance point $z_0$ is~\cite{Carenza:2023nck}
\begin{equation}
    P_{a\gamma}(\infty)=1-e^{- \pi^{2}\left(\frac{L_{\rm res}}{L_{\rm osc}}\right)^{2}}\,,
    \label{eq:estimate}
\end{equation}
where $L_{\rm res}=\sqrt{2\pi/|\partial_{z}\Delta_{\parallel}|}$, evaluated at $z_{0}$, is the resonance length, measuring how fast the plasma frequency varies, and $L_{\rm osc}=2\pi/\Delta_{\rm osc}\simeq \pi/\Delta_{a\gamma}$ is the oscillation length related to the axion-photon mixing at the resonance point. 
While Eq.~\eqref{eq:estimate} is strictly valid only in the relativistic regime, we claim that it can be effectively extended down to the nonrelativistic case, provided the oscillation length $L_\text{osc}$ is computed by  replacing $\Delta_{\rm osc}$ with $\Delta_{\rm osc}^{\rm full}$ [defined in Eq.~\eqref{eq:doscfull}].
Indeed, following Eq.~\eqref{eq:estimate}, a shorter oscillation length would correspond to a larger conversion probability, in agreement with our numerical results.\\
We remark that this constitutes a very naive explanation for the observed behavior, which anyway well captures the physical results also at a quantitative level. Correctly reproducing ultra and mildly relativistic axion conversions, the modified formula loses accuracy only deep down in the nonrelativistic regime, signaling that a more refined analysis should be carried out.
We show Eq.~\eqref{eq:estimate} and its modified counterpart, labeled as $P_{a\gamma}^{\text{full}}(\infty)$, in Fig.~\ref{fig:pcomp}. The relativistic version of Eq.~\eqref{eq:estimate} (dashed green line) correctly reproduces the analytical result presented in Eq.~(3.17) in Ref.~\cite{Carenza:2023nck} , while the counterpart $P_{a\gamma}^{\text{full}}(\infty)$ is in better agreement with numerical simulations, and captures the behavior at low energies. At high energies, the two formulae converge, as expected.
Thus, we conclude that, upon choosing the oscillation length appropriately, Eq.~\eqref{eq:estimate} accurately reproduces the numerical results of Eq.~\eqref{eq:simpleom}. With this example, we point out that adopting a relativistic treatment for mildly to nonrelativistic axions can lead to significant errors in the conversion probability estimates.

\section{Conclusions}
\label{sec:concl}
In the presence of a magnetic field, axions and photons can convert into each other. 
Assuming a constant magnetic field, we improved on the commonly assumed relativistic limit for the axion, and we found expressions for the conversion probability that held on general grounds. We noticed that the amplitude of the probability was correctly retrieved by the relativistic approach; however, in the nonrelativistic limit, the conversion length scale was generally different than what the relativistic limit predicted. This feature may be relevant for axion DM conversion in the neutron star magnetosphere, where the axion could be mildly to nonrelativistic~\cite{Battye:2019aco,Leroy:2019ghm,Witte:2021arp,Battye:2021xvt,Witte:2022cjj,McDonald:2023shx,Tjemsland:2023vvc}, although the formalism should be extended to account for the full phenomenology of the conversion in such extreme environments~\cite{McDonald:2023ohd,Gines:2024ekm}.
In Eq.~\eqref{eq:finalPfull}, we provided a ready-to-use analytical formula describing the conversion probability, in full generality, for a constant magnetic field and plasma frequency.\\
Moreover, we devised a perturbative expansion to be used when the magnetic field or the plasma featured inhomogeneities whose characteristic scale was larger than the typical axion-photon conversion length, given by the inverse of Eq.~\eqref{eq:doscfull}. 
We provided a leading-order accurate analytical formula in Eq.~\eqref{eq:Ppert}, and identified the range of validity of the expansion in Eq.~\eqref{eq:val}. When such a condition is not satisfied, a full numerical analysis should be implemented to determine the correct conversion probability. 
Finally, we considered resonant axion-photon conversions, specializing to the case of a varying plasma frequency and a constant magnetic field. In this setting, we demonstrated how using a naive relativistic formulation can lead to significant underestimations of the conversion probability for mildly to nonrelativistic axions, and we provided a heuristic modification of the usual LZ formula that correctly reproduced the numerical results.
In conclusion, we provided, for the first time, an analytical understanding of the axion-photon mixing valid for all the axion energies, providing simple formulae that can be readily applied for estimates.

\subsection*{Note added}
In the review process, the anonymous referee pointed us to Ref.~\cite{Huang_2018}, which we were not aware of and missed in the literature. Our formula in Eq.~\eqref{eq:finalPfull} agrees with their Eq.~(13) and we agree with their derivation, presenting an extended discussion in Secs.~\ref{sec:basics}-\ref{sec:exact}. In addition to their work, we develop a perturbative treatment of inhomogeneity in Sec.~\ref{sec:pert} and we analyze a case of resonant conversion in Sec.~\ref{sec:LZ}. Moreover, we provide a comparison with previos literature in Appendix~\ref{sec:appa}. We thank the anonymous referee for drawing our attention to such relevant work.

\section*{Acknowledgments}
It is a pleasure to thank Giuseppe Lucente for meaningful discussions and suggestions and for careful proofreading. We thank Samuel J. Witte for clarifications and detailed comments on the text.
C.S. wishes to thank Claudio Gatti and Enrico Nardi for insightful discussions and Giulio Neri for important comments on the revised version of the draft. P.C. thanks  Joshua Eby, M. C. David Marsh and Eike Ravensburg  for discussions in the early stage of the project.
The work of P.C. is supported by the European Research Council under Grant No.~742104 and by the Swedish Research Council (VR) under Grant No.  2018-03641, No. 2019-02337, and No. 2022-04283.
This article is based on work from COST Action COSMIC WISPers CA21106, supported by COST (European Cooperation in Science and Technology).

\section*{Appendix: Comparison with previous literature}\label{sec:appa}
The axion-photon conversion in the presence of a background magnetic field has been studied in different contexts, and different ways to tackle the problem have been developed. In this appendix, we review different methods to solve the coupled system, showing how they connect to our results. As discussed in Sec.~\ref{sec:basics}, the axion-photon system is described by the following coupled equations:
\begin{equation}
    \left(-\partial_t^2 + \nabla^2 + \mathcal{M}^2\right)\phi = 0,
    \label{app:kg}
\end{equation}
where $\phi = (A_\parallel, a)^T$ and the mass matrix is represented by $\mathcal{M}^2$. In the following, we will refer to the diagonal entries of the mass matrix as $\mathcal{M}_1^2$ and $\mathcal{M}_2^2$, while the symmetric off-diagonal terms will be labeled by $\mathcal{M}_\times^2$. The entries of the mass matrix will not be specified further, but they should be thought in analogy to Eq.~\eqref{eq:eom}. At this point, we can generically expand the field $\phi$ in a superposition of plane waves with time- and momentum-dependent coefficients
\begin{equation}
    \int\frac{dp}{2\pi} \tilde {\phi}(p,t) e^{ipz},
\end{equation}
where we focus on a 1D propagation along the $z$-axis. Then, Eq.~\eqref{app:kg} can be written in momentum space, where it becomes:
\begin{equation}
    \left(-\partial_t^2 - p^2 + \mathcal{M}^2\right)\tilde\phi(p,t) = 0.
\end{equation}
Choosing $\tilde\phi(p,t) = \tilde{\phi}(p) e^{-i\omega t}$, we have, for each $p$
\begin{equation}
    \begin{pmatrix}
        \omega^2 -p^2 + \mathcal{M}_1^2& \mathcal{M}_\times^2\\
        \mathcal{M}_\times^2 & \omega^2 -p^2 + \mathcal{M}_2^2
    \end{pmatrix}
    \begin{pmatrix}
        \tilde{A}_\parallel(p)\\
        \tilde{a}(p)
    \end{pmatrix} = 0.
\end{equation}
The system admits a solution only when the determinant is zero. Assuming a static magnetic field, the energy in the conversion process is conserved, and this condition enforces the momentum to be
\begin{equation}
    p^2_{\pm} = \omega^2 + \frac{1}{2}\left(\mathcal{M}_1^2 + \mathcal{M}_2^2\right) \pm \frac{1}{2} \sqrt{\left(\mathcal{M}_1^2 - \mathcal{M}_2^2\right)^2 + 4\mathcal{M}_\times^4}.
    \label{app:ppm}
\end{equation}
This is exactly what we find, in our language, in Eq.~\eqref{eq:ksfull}. It tells us that the most general solution we can look for is
\begin{equation}
    \begin{pmatrix}
        A_\parallel\\
        a
    \end{pmatrix} = 
    \begin{pmatrix}
        \tilde{A}_\parallel^1 e^{-ip_+z} + \tilde{A}_\parallel^2 e^{-ip_-z}\\
        \tilde{a}^1 e^{-ip_+z} + \tilde{a}^2 e^{-ip_-z}        
    \end{pmatrix} e^{i\omega t},
    \label{app:ev}
\end{equation}
which is exactly what we find in Eq.~\eqref{eq:evol_full}. This formalism has been employed in several works, see, e.g., Ref.~\cite{MAIANI1986359}. The difference with Ref.~\cite{MAIANI1986359} is that they consider a fixed $p$ and varying $\omega$ and they focus on the effects on light propagation. Instead, for us, $\omega$ is fixed, and $p$ varies. To draw a connection with our formalism, we will show that this leads to the same conversion probability that we find in Eq.~\eqref{eq:finalPfull}. To do so, we fix the initial conditions
\begin{align}
    &A_\parallel(t = 0, z = 0)  = 1\Longrightarrow \tilde{A}_\parallel^2 = 1- \tilde{A}_\parallel^1\\
    &a (t = 0, z = 0) = 0 \Longrightarrow \tilde{a}^2 = - \tilde{a}^1;
\end{align}
in other words, our initial state has one photon and no axion.
Then, we have
\begin{equation}
    \left(\omega^2 + \partial_z^2 + \mathcal{M}^2\right)
    \begin{pmatrix}
        \tilde{A}_\parallel^1 e^{-ip_+z} + (1-\tilde{A}_\parallel^1) e^{-ip_-z}\\
        \tilde{a}^1 (e^{-ip_+z} - e^{-ip_-z})
    \end{pmatrix} = 0.
\end{equation}
Solving the coupled system yields
\begin{equation}
    \tilde{A}_\parallel^1 = \frac{\omega^2-p_-^2 + \mathcal{M}_1^2}{p_+^2 - p_-^2}\qquad \tilde{a}^1 = \frac{\mathcal{M}_\times^2}{\sqrt{\left(\mathcal{M}_1^2 - \mathcal{M}_2^2\right)^2 + 4\mathcal{M}_\times^4}}.
\end{equation}
At this point, we have
\begin{equation}
        \begin{pmatrix}
        A_\parallel\\
        a
    \end{pmatrix} = 
    \begin{pmatrix}
        \frac{\omega^2-p_-^2 + \mathcal{M}_1^2}{p_+^2 - p_-^2} e^{-ip_+z}- \frac{\omega^2-p_+^2 + \mathcal{M}_1^2}{p_+^2 - p_-^2} e^{-ip_-z}\\
         \frac{\mathcal{M}_\times^2}{\sqrt{\left(\mathcal{M}_1^2 - \mathcal{M}_2^2\right)^2 + 4\mathcal{M}_\times^4}} \left(e^{-ip_+z} - e^{-ip_-z}\right)        
    \end{pmatrix} e^{i\omega t},
    \label{app:fullmat}
\end{equation}
from which the conversion probability follows very easily,
\begin{equation}
    P_{a\gamma} = \left|a\right|^2 = \frac{4\mathcal{M}_\times^4}{\left(\mathcal{M}_1^2 - \mathcal{M}_2^2\right)^2 + 4\mathcal{M}_\times^4}\sin^2\left(\frac{1}{2}\left(p_+ - p_-\right)z\right),
\end{equation}
and perfectly agrees with the probability we found in Eq.~\eqref{eq:finalPfull}. Again, we see that the axion and the photon cannot be considered as plane waves with a definite value of the momentum, rather, they are described by the specific superposition of plane waves in Eq.~\eqref{app:ev}. As our analysis highlights, the underlying reason is that axions and photons are not propagation eigenstates, and, therefore, we do not expect them to propagate following a simple plane wave behavior.\\
As the strength of the mixing term incorporating the magnetic field decreases, however, we expect the plane wave description to be more reliable. Indeed, in the limit of zero magnetic field, axions and photons are decoupled, and so they can rightfully be described by plane waves with fixed values of momentum. Therefore, let us assume that $\mathcal{M}_\times^2 \ll |\mathcal{M}_1^2 - \mathcal{M}_2^2|$. Then, we can expand Eq.~\eqref{app:ppm} as 
\begin{equation}
    p^2_{\pm} = \omega^2 + \frac{1}{2}\left(\mathcal{M}_1^2 + \mathcal{M}_2^2\right) \pm \frac{1}{2}\left(\mathcal{M}_1^2 - \mathcal{M}_2^2\right) \pm \frac{\mathcal{M}_\times^4}{\mathcal{M}_1^2 - \mathcal{M}_2^2},
\end{equation}
or, explicitly,
\begin{align}
    &p^2_+ = \omega^2 + \mathcal{M}_1^2 + \frac{\mathcal{M}_\times^4}{\mathcal{M}_1^2 - \mathcal{M}_2^2}\\
    &p^2_- = \omega^2 + \mathcal{M}_2^2 - \frac{\mathcal{M}_\times^4}{\mathcal{M}_1^2 - \mathcal{M}_2^2},
\end{align}
where we assumed $\mathcal{M}_1^2 - \mathcal{M}_2^2 >0$, without loss of generality.
Substituting this into the coefficients of the first row of Eq.~\eqref{app:fullmat} gives
\begin{align}
    &\tilde{A}_\parallel^1 = \frac{(\mathcal{M}_1^2 - \mathcal{M}_2^2)^2 + \mathcal{M}_\times^4}{(\mathcal{M}_1^2 - \mathcal{M}_2^2)^2 + 2\mathcal{M}_\times^4},\\
    &\tilde{A}_\parallel^2 = \frac{ \mathcal{M}_\times^4}{(\mathcal{M}_1^2 - \mathcal{M}_2^2)^2 + 2\mathcal{M}_\times^4}.
\end{align}
Therefore, in the $\mathcal{M}_\times^2 \ll |\mathcal{M}_1^2 - \mathcal{M}_2^2|$ limit, the plane wave ansatz represents a good approximation. However, when the magnetic field is strong or, alternatively, when we consider resonance, as in Sec.~\ref{sec:LZ}, the plane wave ansatz fails, even if the plasma is homogeneous and the magnetic field is constant.
In other words, the plane wave ansatz works only in the limit $\theta \ll 1$, where $\theta$ is defined by Eq.~\eqref{eq:tantheta}, while our formalism works even when extrapolated to large $\theta$, as it happens close to resonances.\\
To gain a deeper understanding, let us consider an ultrarelativistic axion, with energy $\omega$ so large, compared to its mass, that we can effectively set $m_a \sim 0$. For simplicity, let us relabel the entries of the mass matrix as 
\begin{equation}
    \mathcal{M}_1^2 \rightarrow Q_\parallel, \qquad \mathcal{M}_2^2 \rightarrow Q_a, \qquad \mathcal{M}_\times^2 \rightarrow Q_{a\gamma}, 
\end{equation}
in accordance with the analysis in the main text.
Then, let us assume that also $Q_\parallel \sim 0$. Under these conditions, the relativistic prescription for the oscillation length dictates
\begin{equation}
    \Delta_\text{osc} \sim \frac{Q_{a\gamma}}{\omega},
    \label{app:doscrel}
\end{equation}
where we used Eqs.~\eqref{eq:doscrel} and \eqref{eq:di}. The complete approach developed in Sec.~\ref{sec:exact}, instead, yields
\begin{align}
    \Delta_\text{osc}^\text{full}&\sim \sqrt{\omega^2 + Q_{a\gamma}} - \sqrt{\omega^2 - Q_{a\gamma}}
    \sim \nonumber\\
    &\sim\begin{cases}
        \frac{Q_{a\gamma}}{\omega}\qquad &\text{if}\quad \omega \gg g_{a\gamma}B_0\\
        \newline
        \sqrt{2}g_{a\gamma}B_0 \qquad &\text{if}\quad \omega = g_{a\gamma}B_0
    \end{cases}
    \label{app:doscfull}
\end{align}
substituting in Eq.~\eqref{eq:doscfull} and using  Eq.~\eqref{eq:invac} for $Q_{a\gamma}$. Therefore, even if we focus on an ultrarelativistic axion, with $\omega \gg m_a$, the relativistic approach can produce inaccurate results if $\omega \sim g_{a\gamma} B_0$. Indeed, in Sec.~\ref{sec:exact} we clearly specified that the relativistic approximation is really the statement $\omega \gg M$, where $M$ is a placeholder for $|Q'_\parallel|^{1/2}, |Q_a'|^{1/2}$. In this case, when $\omega \sim g_{a\gamma} B_0$, the relativistic condition is not satisfied.\\
\begin{figure}[t!]
    \centering
    \vspace{0ex}%
    
    \includegraphics[width=0.43\textwidth]{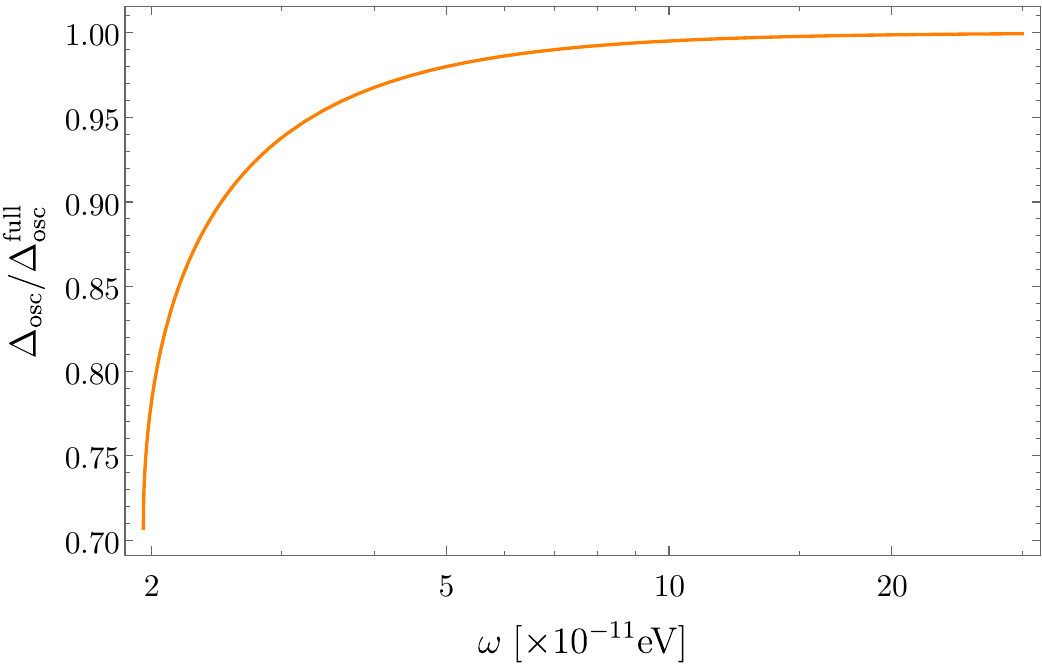}
    \includegraphics[width=0.43\textwidth]{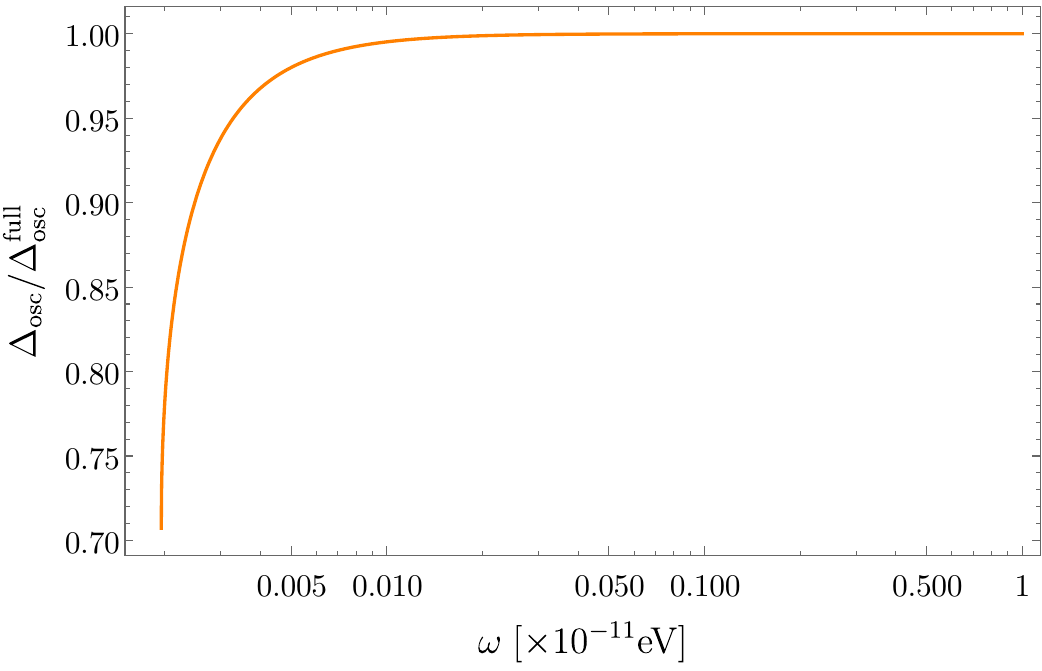}
    \caption{Plot of the ratio between the relativistic oscillation length [Eq.~\eqref{app:doscrel}] and the full oscillation length [Eq.~\eqref{app:doscfull}], as a function of the initial axion energy $\omega$. In both panels, we assume $Q_a, Q_\parallel \sim 0$ (see main text). Moreover, we fix the axion-photon coupling to $g_{a\gamma} = 10^{-10}~\text{GeV}^{-1}$. In the top panel, we consider a magnetic field intensity of $B_0 = 10^{10}~\text{G}$, while the bottom panel shows the result for $B_0 = 10^{7}~\text{G}$. }
    \label{fig:appratio}
\end{figure}
\begin{figure}[t!]
    \centering
    \vspace{0ex}%
    \includegraphics[width=0.43\textwidth]{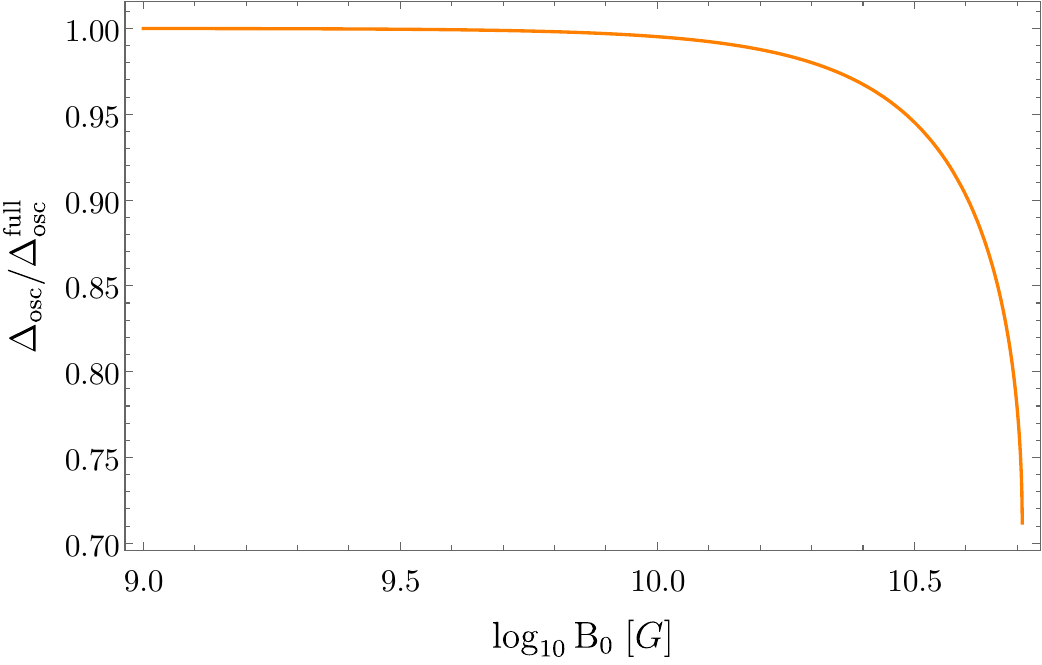}
    \caption{Plot of the ratio between the relativistic oscillation length [Eq.~\eqref{app:doscrel}] and the full oscillation length [Eq.~\eqref{app:doscfull}], as a function of the magnetic field intensity $B_0$. In the plot, we assume $Q_a, Q_\parallel \sim 0$ (see main text). Moreover, we fix the axion-photon coupling to $g_{a\gamma} = 10^{-10}~\text{GeV}^{-1}$ and the axion energy to $\omega = 10^{-10}~\text{eV}$. }
    \label{fig:appratioB}
\end{figure}
We plot the ratio $\Delta_\text{osc}/ \Delta_\text{osc}^\text{full}$ in Fig.~\ref{fig:appratio}. In the top panel, we set $g_{a\gamma} = 10^{-10}~\text{GeV}^{-1} $ and $B_0 = 10^{10}~\text{G}$, while the bottom panel shows the same value of $g_{a\gamma}$, but $B_0 = 10^{7}~\text{G}$. 
We conveniently normalize $\omega$ to an arbitrary energy scale of $10^{-11}~\text{eV}$ for better readability.
As expected, the behavior of the ratio $\Delta_\text{osc}/ \Delta_\text{osc}^\text{full}$ smoothly interpolates from $1/\sqrt{2}$ to 1, in agreement with the theoretical estimates in Eqs.~\eqref{app:doscrel} and \eqref{app:doscfull}. In Fig.~\ref{fig:appratioB}, instead, we focus on an axion with $\omega = 10^{-10}~\text{eV}$ and $g_{a\gamma} = 10^{-10}~\text{GeV}^{-1} $, and we plot the ratio $\Delta_\text{osc}/ \Delta_\text{osc}^\text{full}$ as a function of the magnetic field. As the magnetic field increases, $Q_{a\gamma}$ grows, until the condition $\omega = g_{a\gamma} B_0$ is saturated. In this case, the ratio $\Delta_\text{osc}/ \Delta_\text{osc}^\text{full}$ decreases from 1, when $B_0 \ll \omega/g_{a\gamma}$, to $1/\sqrt{2}$, when $B_0 \sim \omega/g_{a\gamma}$ and the relativistic approximation cannot be applied anymore.
With this discussion at hand, we can also understand the differences with other papers, where the axion-photon conversion is analyzed from a quantum field theory perspective, in terms of Feynman diagrams~\cite{Sikivie:1983, Sikivie:1985, Pshirkov_2009}. In particular, they have to calculate the matrix element
\begin{equation}
    \left.\langle \gamma(p)|\frac{1}{4} g_{a\gamma} a F_{\mu\nu} \tilde{F}^{\mu\nu}|a(q)\rangle\right|_{B_\text{0}},
\end{equation}
where $\gamma(p)$ and $a(q)$ are the coefficients of the plane wave expansion of the incoming axion and the outgoing photon, with specific momentum $q$ and $p$, respectively. The computation of this matrix element assumes that the axion and photon behave as plane waves with a specific momentum, which is only true perturbatively. In other words, saying that axions and photons have a definite momentum is only strictly true when the magnetic field is not present, for the system~\eqref{app:kg} becomes decoupled and Eq.~\eqref{app:ppm} truly defines the momenta of the axion and the photon. In light of this observation, it would be worth reviewing the resonant axion-photon conversion in neutron star magnetospheres, exploring the regime of validity of the commonly assumed plane wave ansatz and Wentzel-Kramers-Brillouin approximation~\cite{Leroy:2019ghm, Witte:2021arp}. We leave this analysis to future work.  
\bibliography{biblio.bib}

\end{document}